\documentclass[12pt]{article}
\usepackage[margin=0.75in,top=0.6in]{geometry}
\usepackage[utf8]{inputenc}
\usepackage{amsmath,amsfonts,amssymb}
\usepackage{algorithm}
\usepackage{algpseudocode}
\usepackage{graphicx}
\usepackage{subfig}
\usepackage{float}
\usepackage{multirow} 
\usepackage{array}  
\usepackage{booktabs}  
\usepackage{siunitx}   
\usepackage{caption}
\usepackage{hyperref}
\usepackage{authblk}
\date{}
\title{Simulation of Stochastic Discrete Dislocation Dynamics in Ductile Vs Brittle Materials}

\author[1]{Santosh Chhetri}
\author[2]{Maryam Naghibolhosseini}
\author[1,3]{Mohsen Zayernouri}
\affil[1]{Department of Mechanical Engineering, Michigan State University}
\affil[2]{Department of Communicative Sciences and Disorders, Michigan State University}
\affil[3]{Department of Statistics and Probability, Michigan State University}

\begin{document}

\maketitle

\begin{abstract}
    Defects are inevitable during the manufacturing processes of materials. Presence of these defects and their dynamics significantly influence the responses of materials. A thorough understanding of dislocation dynamics of different types of materials under various conditions is essential for analyzing the performance of the materials. Ductility of a material is directly related with the movement and rearrangement of dislocations under applied load. In this work, we look into the dynamics of dislocations in ductile and brittle materials using simplified two dimensional discrete dislocation dynamics (2D-DDD) simulation. We consider Aluminium (Al) and Tungsten (W) as representative examples of ductile and brittle materials respectively. We study the velocity distribution, strain field, dislocation count, and junction formation during interactions of the dislocations within the domain. Furthermore, we study the probability densities of dislocation motion for both materials. In mesoscale, moving dislocations can be considered as particle diffusionm, which are often stochastic and super-diffusive. Classical diffusion models fail to account for these phenomena and the long-range interactions of dislocations. Therefore, we propose the nonlocal transport model for the probability density and obtained the parameters of nonlocal operators using a machine learning framework.
\end{abstract}

{
\hspace{10mm}\noindent \textbf{Keywords:} 
\footnotesize{Discrete Dislocation Dynamics, Ductile Materials, Brittle Materials, Nonlocal Models} \\
}

\section{Introduction}

Crystalline materials are solids characterized by a highly ordered and repeating arrangement of atoms in a three-dimensional structure. This orderly pattern extends throughout the entire structure, forming a crystal lattice that is consistent and repetitive in all directions. They are used in a wide array of applications due to their unique and specific properties that arise from their ordered atomic structures. Dislocations within the crystalline materials are the fundamental crystallographic defects within their structure. Presence and various activities of dislocations define the various mechanical properties of the materials and responsible mechanical responses such as ductility or brittleness of the material \cite{bulatov2006computer}.  Experimentally it has been observed that dislocations move within the domain with the velocity comparable to the speed of sound \cite{gumbsch1999dislocations, gurrutxaga2021mechanics}, resulting in permanent deformation of the materials \cite{bulatov1998connecting}. Under the influence of external load, dislocations jump from original shear plane to new one \cite{li2002dislocation, de2021atomistic}, accumulation of these activities leads to the initiation of fatigue cracks \cite{zhu2004atomistic}, and material failure \cite{tanaka1981dislocation}. Dislocations in ductile materials can move relatively freely, allowing the material to deform under stress \cite{pang2014dislocation}, conversely brittle materials have fewer available slip systems, and the movement of dislocations is much more restricted \cite{li2019dislocation}. 

Long-range interactions of dislocations results in intermittency, strain bursts and power law scaling in velocity distribution \cite{miguel2001intermittent}. Such findings are indicative of anomalous diffusion processes, where the mean square displacement (MSD) exhibits a nonlinear behavior and follows a power-law relationship, unlike the linear relationship observed in traditional diffusion through Brownian motion. Phase-Field Model (PFM) has emerged as a powerful mathematical tool for capturing material damage \cite{barros2021integrated} and failure prediction\cite{de2021data, de2024thermo, kc2024thermo, kc2024multi} . It offers a robust framework for simulating complex fracture phenomena, such as crack initiation, propagation, and coalescence, across a wide range of materials. Despite its versatility in modeling macroscopic failure mechanisms, it fails to provide the detailed dislocation dynamics for accurately predicting the material behaviour, particularly in cases which involves plastic deformation. Out of all the available computer simulation methods, DDD simulation is an appropriate method to observe such phenomenon at the micron scale \cite{davoudi2018dislocation, arsenlis2007enabling}. From early phenomenological models \cite{holt1970dislocation, walgraef1985dislocation} to stochastic approaches\cite{hahner1996theory, kapetanou2015statistical}, DDD simulation provided the valuable insights for materials behaviour. Ductile materials can bear significant plastic deformation before fracture, while brittle materials shows little to no plastic deformation.

Classical differential models fail to incorporate the size effect, complex core structures \cite{sandfeld2011continuum, liu2017atomically}. Nonlocal models can be used as an alternative to the traditional differential models by allowing for discontinuities and incorporating long-range interactions inherently within an integral formulation. These characteristics are appealing for addressing challenges in convection-diffusion \cite{ du2014nonlocal, d2017nonlocal}, anomalous materials \cite{suzuki2021thermodynamically}, turbulent flows \cite{seyedi2022data, samiee2020fractional, akhavan2021data, samiee2022tempered}, subsurface dispersion \cite{schumer2001eulerian},and  heterogeneous media \cite{du2016multiscale}. The peridynamic theory \cite{silling2000reformulation} emerged as a nonlocal substitute for the traditional continuum mechanics of solids, finding utility particularly in fracture scenarios involving discontinuities \cite{silling2003dynamic, silling2014peridynamic}. In recent years, there has been substantial discourse regarding the formulization of nonlocal models into a nonlocal vector calculus \cite{du2012analysis, du2013nonlocal} alongside progress toward integrating nonlocal/fractional models \cite{gulian2021unified, d2013fractional, tzelepis2023polyurea, khoshnevis2024stochasticgeneralizedorderconstitutivemodeling}.

The rise of Machine Learning (ML) techniques has grown across various fields of materials science and physics \cite{steinberger2019machine, bertin2020frontiers, hiemer2023relating}, where learning algorithms are employed to improve comprehension of physics, estimate model parameters, or develop resilient surrogates using high-fidelity data. Recently, there has been a surge in interest in employing data-driven methods for dislocation dynamics. In \cite{salmenjoki2018machine} , authors explored the predictability of plastic deformation in micron-scale crystalline solids using ML techniques.

Alternative ML methods have also advanced in the domain of understanding physics via Partial Differential Equations (PDEs). We recognize the advancements brought by Physics-Informed Neural Networks (PINNs)\cite{raissi2019physics}, in enhancing deep neural networks with physics-based constraints, as well as the techniques for discovering PDEs using candidate terms and operators \cite{bakarji2021data, lee2020coarse, rudy2017data, supekar2023learning}. Furthermore, research has explored the application of ML techniques to uncover differential equations that govern the evolution of Probability Density Functions (PDFs). For instance, in article \cite{bakarji2021data}, authors employed Kernel Density Estimation (KDE) to derive probability densities from discrete stochastic processes. This fundamental approach of inferring PDF equations from stochastic discrete data was further utilized by \cite{maltba2022autonomous} in their investigation of non-local closure terms. Additionally, in article \cite{brennan2018data}, authors have further investigated the data-driven discovery of closure terms for PDF equations.

In the current research, we compare the velocity field, accumulated plastic strain along with the junction formation and dislocation density. In addition to that, we employ two-dimensional DDD simulations to produce probability distribution functions for both brittle and ductile materials based on displaced dislocation positions gathered from multiple DDD instances. This method allows us to directly observe the dynamic behavior of dislocation positions and analyze the behaviour for different materials. We convert the individual particle dynamics into a continuous PDF using an Adaptive Kernel Density Estimation (AKDE) technique, resulting in a sequence of PDFs representing dislocation positions over time. We introduce a nonlocal model that utilizes a kernel-based integral operator to describe the evolution of these PDFs as the continuum approximation of the stochastic process, and we use the ML framework developed and used in our previous work \cite{de2023machine, chhetri2023comparative}, to obtain the parameters of this nonlocal kenrel by analyzing the PDF generated from DDD data.

We summarize our main contributions below.
\begin{itemize}

    \item We study the key parameters extracted from the DDD simulation and look into the statistics for analysis of dislocation behaviour in ductile and brittle materials.
    
    \item We obtain the PDF from DDD simulation data for both materials and highlight the differences in behaviour of dislocations under different loading conditions. 

    \item We derive the parameters of the nonlocal power-law kernel, which include the $\alpha$, $\delta$ and a linear coefficient, for both ductile and brittle materials.

\end{itemize}

\section{Two-dimensional discrete dislocation dynamics}
DDD simulation offers a dependable framework for studying plastic deformation and fracture phenomena that take place at the micron level due to the interactions of numerous dislocations. Observing such phenomena using conventional linear elasticity or atomistic approaches would pose significant challenges. A simplified 2D-DDD simulation model serves as a suitable method for extracting crucial parameters such as velocity distribution, stress, and plastic strain \cite{salmenjoki2018machine}.

We placed $N$ straight edge dislocations with a Burgers vector direction $\bf{\textit{$b$}} = \pm b$, aligned along the $z$ axis, randomly within a predefined square area of size $L$. The dislocations are permitted to move exclusively along the $x$ direction; any movement in the $y$ direction, such as climbing or jumping, is constrained.

Dislocations within a domain are under the influence of long-range interaction stress, as well as any applied load externally ($\bf{\sigma_{ext}}$). In contrast to the classical diffusion, long-range interactions also have a significant effect on the mobility of dislocation. A dislocation at any given location experiences a stress ($\sigma_i({r})$) due to the presence of another dislocation at distance $x$ and $y$ and  can be calculated using\cite{anderson2017theory}, 

\begin{equation}
\sigma_i({r}) = \frac{\mu b}{2 \pi (1-\nu)} \frac{x(x^2 - y^2)}{(x^2 + y^2)^2},
\label{nlml_Eq:continuity_equation}
\end{equation}

where $x$ and $y$ is the distance between the edge dislocation along the $x$ and $y$ axis, $\mu$ is the bulk shear modulus, $b$ is the Burger's  and $\nu$ is the Poisson's ratio.

We selected a representative unit from the material's bulk and implemented Periodic Boundary Conditions (PBC) to maintain a consistent number of dislocations within the simulation area, assuming that the surfaces are distant enough from any free surfaces. To account for the influence of replicated surfaces and interactions over long distances, we employ the subsequent formula to calculate the stress field \cite{van1995discrete}.

\begin{equation}
\sigma_i({r}) = \frac{\mu b}{2 (1-\nu)} \frac{1}{L} \frac{\sin(X) \left[\cosh(Y) - \cos(X) -Y\sinh(Y)\right]}{\left[\cosh(Y)-\cos(X)\right]^2},
\label{nlml_Eq:periodic_equation}
\end{equation}

\noindent where $X = \frac{2 \pi x}{L}$ and $Y = \frac{2 \pi y}{L}$.

The Peach-Koehler equation (Eq \ref{nlml_Eq:glide equation}) enables the calculation of the total gliding force (known as the Peach-Koehler force) per unit length acting on the $i^{th}$ dislocation due to internal interactions in the absence of any external load \cite{gulluoglu1989dislocation}.

\begin{equation}
\frac{F^{g}_i}{L} = \sum_{i\neq j}\frac{\mu b_i b_j}{2 \pi (1-\nu)}\frac{x(x^2 - y^2)}{(x^2 + y^2)^2}
\label{nlml_Eq:glide equation}
\end{equation}

Considering the motion of dislocation to be a over damped motion, and with the restriction of any movement along the $y$ direction, constitutive governing equation of motion of a $i^{th}$ dislocation along the glide the plane is:
\begin{equation}
\eta \frac{d x_i}{dt} = b_i \left(\sum_{m\neq i}^{N} \sigma_i (r_m - r_i) + \sigma_{ext} \right),
\label{nlml_Eq:mobility}
\end{equation}

where \noindent $\eta$ is the mobility parameter.

All the stresses mentioned in this work are the shear stress $\tau_{xy}$, such that the resultant stress can be considered as the gliding stress along the glide plane. This equation can be solved using Forward-Euler scheme. The total accumulated plastic strain of the simulation domain due to $N$ number of dislocation shifts can be obtained using the Orowan's formula \cite{hull2001introduction}, 
\begin{equation}
\gamma = \frac{1}{L^2} \sum_{i = 1}^{N} b_i \Delta x_i .
\label{strain}
\end{equation}

In actual scenario, the mutual interaction stress limits two dislocations near a core region, and the constitutive law from linear elasticity becomes invalid and producing an unphysical stress field. Annihilation, multiplication and the junction formation are natural phenomenon that occurs during the movement of dislocations within the domain. It becomes crucial to acknowledge these effects, and implement them in the simulation to obtain more realistic results. 

If two dislocations of opposite sign come closer, they annihilate each other, Which can not be modelled using simplified 2D-DDD. In order to implement, when two opposite sign dislocations are within a predefined core distance $d_a$, we consider them to be annihilated and accordingly removed from the simulation.

Multiplication is the process when existing dislocations within the materials generate new dislocations. This phenomenon is also missing in the simplified 2D-DDD simulation. We followed the similar approach used in \cite{van1995discrete} to implement the multiplication effect in our simulation. We distribute a total of ($N_s$) sources randomly across the domain, and at every iteration, we monitor the stress levels. Should the stress resulting from interactions at a source surpass the critical stress ($\tau_c$) for a duration equal to or greater than the critical time ($t_c$), a new pair of dislocations is created on either side of the source, with a separation distance of $L_{nuc}$. The critical stress at each source is determined by:
\begin{equation}
\tau_c = \frac{\mu b}{2 \pi (1-\nu)} \frac{1}{L_{nuc}}.
\end{equation}

\begin{figure}[H]
	\centering
	\subfloat[Annihilation.]{\includegraphics[width=0.4\textwidth]{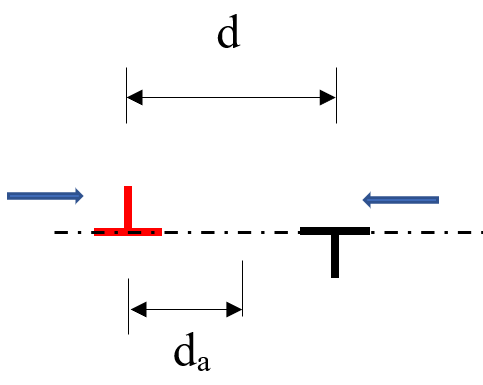}}
    \subfloat[Dislocation generation.]{\includegraphics[width=0.4\textwidth]{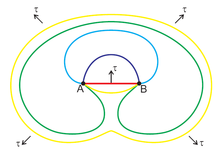}}
	\caption{Illustrative diagram for (a) annihilation and (b) multiplication}
	\label{fig:Ann-Mul}
\end{figure}

Junction formation is also a natural phenomenon, which is a very critical aspect of dislocation behaviour and significantly influences the mechanical properties of materials. When two dislocations come together, sometime they form a stable or semi-stable junction. They become immobile and act as obstacles to other moving dislocations. In order to mimic this, following the similar approach used in \cite{keralavarma2016strain} for the annihilation process, when two dislocations of same sign come within the pre-defined distance $d_j$, we lock them and prevent them from moving. At this stage the mutual stress between these pairs is set to be zero. Then, we check the stress level at every time steps, if the resultant stress level is higher than the breaking stress (which is equal to the nucleation stress $\tau_c$) for a duration equal to or greater than the critical time ($t_c$), the junction is broken and they are free to move again. Another possibility for breaking these junction is when any of the dislocations gets annihilated then remaining dislocation is free to move again. 

\begin{figure}[H]
    \centering
    \includegraphics[width=0.45\textwidth,height=0.4\textheight,keepaspectratio]{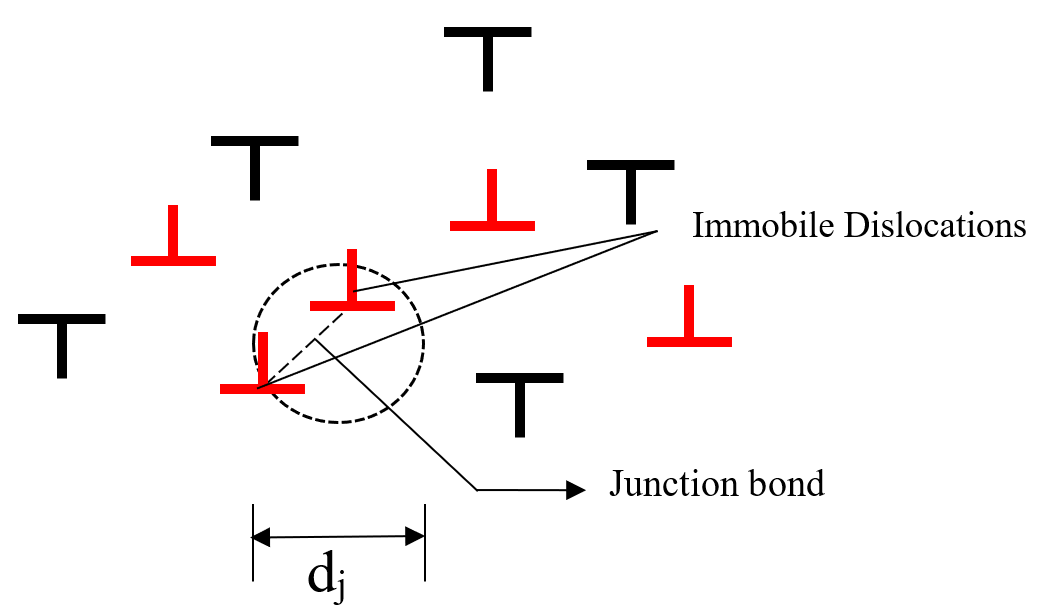}
    \caption{Illustrative figure for junction formation and immobile dislocations}
    \label{fig:junctions}
\end{figure}

\section{Representative Example: Crystal Under Creep}

Failure of materials can be directly related with the dynamical behaviour of the dislocations interactions. Usually metal possesses very high number of dislocations density in the order of $10^{9} cm^{-2}$ \cite{gracca2008determination}. Dislocations become highly dense surpassing this number during the time of failure. Here, in this study we selected Aluminium as a representative case of ductile materials and Tungsten for the brittle materials. We defined a square domain of $L = 300  b$, where we distributed $N= 500$ dislocations randomly over the domain. We set the annihilation distance to be $d_a =  3b$, and junction formation distance to be $d_j = 2 b$. We also distributed $N_s = 50$ dislocations sources over the domain with mean critical distance $\bar{L} = 50  b$, and critical nucleation time $\bar{t} = 10dt$, where we set $dt = 10^{-14} $ seconds.

We perform a creep test on three crystals of each materials with the following loading conditions, 
\begin{itemize}
    \item $\bf{Case 1}$ - $\sigma_{ext} = 100$ MPa; without multiplication
    \item $\bf{Case 2}$ - $\sigma_{ext} = 100$ MPa; with multiplication
    \item $\bf{Case 3}$ - $\sigma_{ext} = 200$ MPa; with multiplication

    Case 1 will be used for the baseline comparison only.
\end{itemize}

In Case 1, we examine areas within the mechanical component where there are no imperfections, impurities, or microcracks. In these regions, the dislocations already present in the material simply move without multiplying, occasionally colliding and annihilating each other. Consequently, Case 1 characterizes zones with reduced internal stresses, plastic deformation, less pronounced dislocation movement, and the absence of noticeable rapid failure phenomena.

On the contrary, Cases 2 and 3 feature sources for multiplication introduced in a conceptual manner, depicting areas within a component where we typically anticipate greater degradation. These regions are characterized by the presence of voids, impurities, microcracks, and uneven surfaces, which naturally serve as generators of dislocations and are commonly associated with zones prone to failure. Thus, Cases 2 and 3 provide insight into the behavior near locations that are likely to induce failure.

\begin{figure}[H]
    \centering
    \subfloat[Initial configuration.]{\includegraphics[width = 0.48\textwidth]{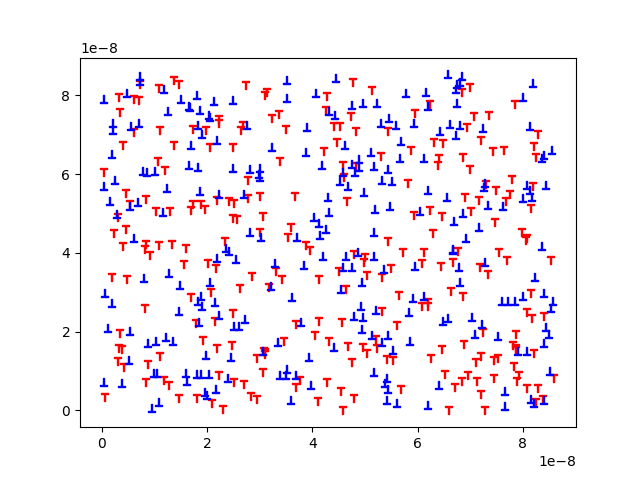}}
    \subfloat[Meta stable configuration.]{\includegraphics[width = 0.48\textwidth]{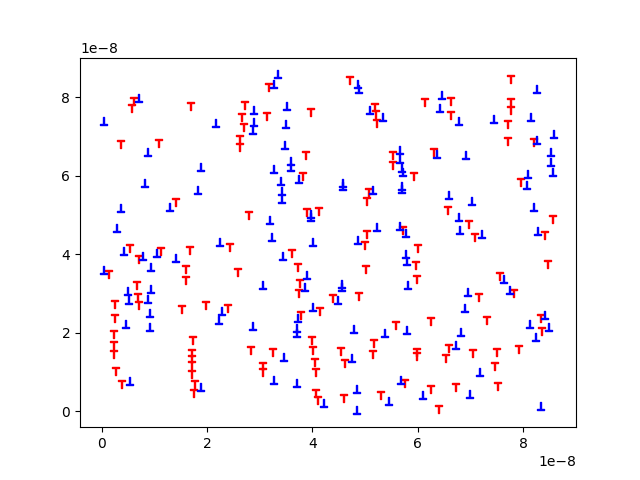}}
    \caption{Dislocation density (a) before relaxation (b) after relaxation.}
    \label{fig:configuration}
\end{figure}

In the figure \ref{fig:configuration}, initial configurations plot shows the highly dense dislocations. Such structure are physically unstable. We let the system relax without applying any external load to obtain the metastable configuration.

\begin{figure}[H]
    \centering
    \subfloat[Number of dislocation.]{\includegraphics[width = 0.3\textwidth]{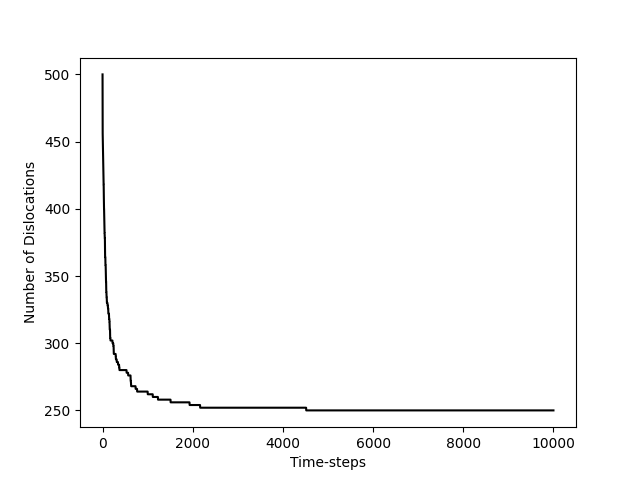}}
    \subfloat[Collective velocity.]{\includegraphics[width = 0.3\textwidth]{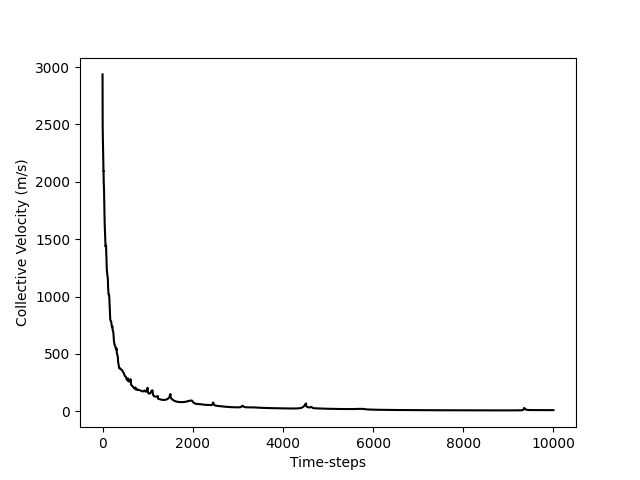}}
    \subfloat[Plastic strain]{\includegraphics[width = 0.3\textwidth]{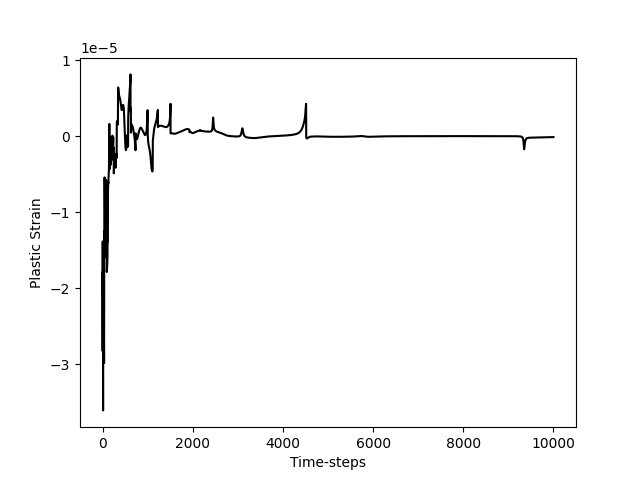}}
    \caption{Time-series plots of dislocation count, average velocity, and plastic strain over the duration of relaxation.}
    \label{fig:stable}
\end{figure}

In the figure \ref{fig:stable}, we plot the number of dilocations, collective velocity and plastic strain over 10000 time steps of relaxation. It can be seen that after about 5000 time steps there is almost no fluctuation in dislocation count, velocity and plastic strain. Thus, it can be assured that the relaxation applied before the external load has achieved the metastable configuration, and system is relaxed.

The DDD simulations are executed using a custom Python script on Intel Xeon Gold 6148 CPUs running at 2.40GHz. Initially, the system is allowed to relax for 10,000 time-steps, each with a duration of \(\Delta t = 1\ dt\), without any external stress applied. During this relaxation phase, significant dislocation activity and annihilation events occur, leading to a metastable state with approximately half the initial number of dislocations. After this, an external stress \(\sigma_{ext}\) is applied, and the simulation continues with a reduced time-step of \(\Delta t = 0.01\ dt\) until reaching a total of 25,000 steps for all scenarios.

\begin{figure}[H]
    
    \subfloat{{\includegraphics[width=9 cm]{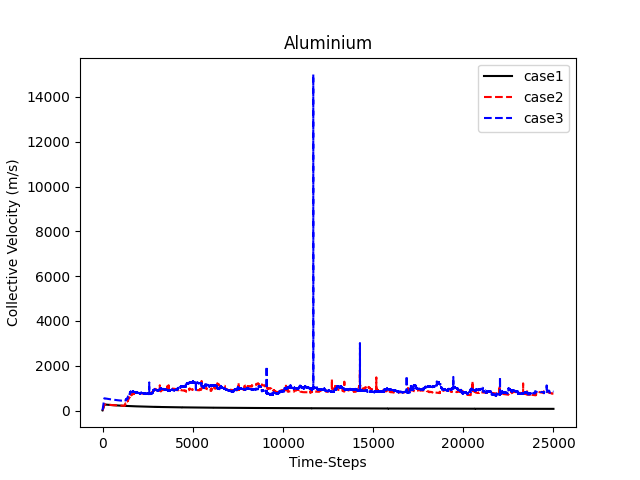} }}
    \subfloat{{\includegraphics[width= 9 cm]{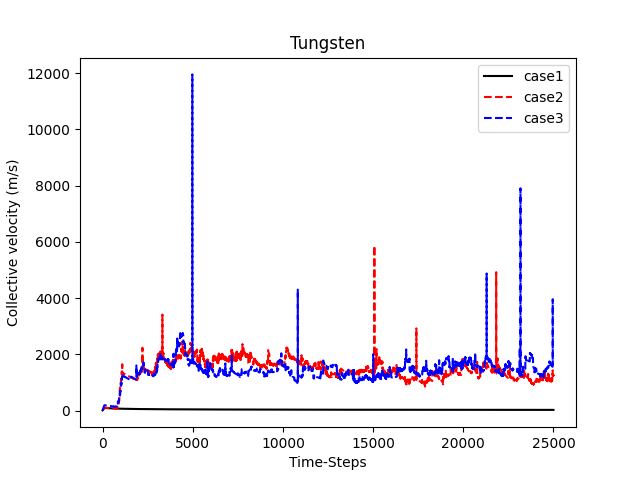} }}
  
    \caption{Collective velocity of dislocation}%
    \label{fig:collective velocity}

\end{figure}

The intermittency in the velocity of dislocations refers to non-continuous, sporadic movement of dislocations through the crystal lattice. This is due to the complex interaction between the dislocation with other dislocations and obstacles, sudden creation of new dislocations from the nucleation source. The link between the intermittency in dislocation velocity, strain bursts, and the failure of materials is critical. High dislocation densities and the resulting strain localization can lead to the initiation and propagation of cracks, ultimately leading to material failure. In the figure \ref{fig:collective velocity}, we observe higher number of intermittency for the Tungsten with higher average velocity. For instance, dislocations in Aluminium for in $\bf{case\ 1}$ are moving with average velocity of $ ~ 115$ $m/s$, while they are moving at the slower pace of $~ 40 $ $m/s$ in tungsten. However, for $\bf{case\ 2}$ and $\bf{case\ 3}$, dislocations in Tungsten move with higher velocity than Aluminium. Table \ref{tab:velocity table} contains the average velocity of dislocations in both crystals.

\begin{table}[htbp]
    \centering
     \caption{Average Velocity (m/s)}
    \begin{tabular}{lccc}
        \toprule
        & \bf{Case 1} & \bf{Case 2} & \bf{Case 3} \\
        \bf{Aluminium} & ${\sim 115}$ & $\sim 845$ &$\sim 900$ \\
        \bf{Tungsten} & ${\sim 40}$ & $\sim 1443$ &$\sim 1404$ \\
        \bottomrule
    \end{tabular}
    \label{tab:velocity table}
\end{table}

\begin{figure}[H]
    
    \subfloat{{\includegraphics[width=9 cm]{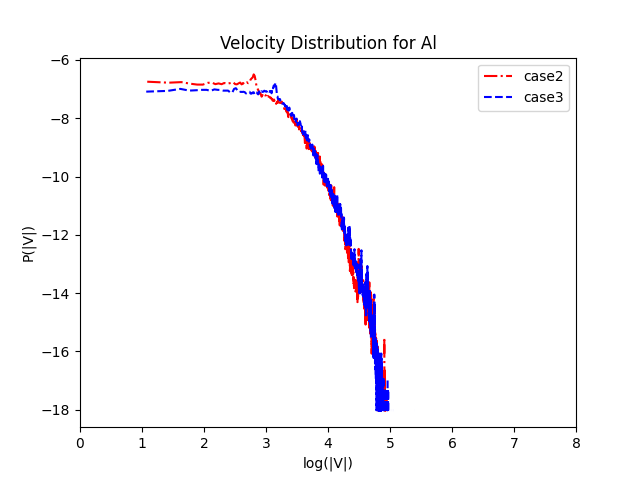} }}
    \subfloat{{\includegraphics[width= 9 cm]{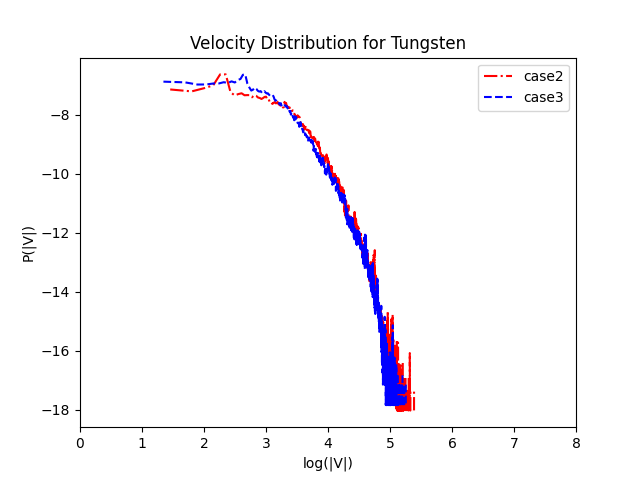} }}
  
    \caption{Probability distribution of the velocity $(P \mid V \mid )$} %
    \label{fig:PDF velocity}

\end{figure}

In order to qualitatively analyze the intermittency, we plot the probability distribution of the velocity. Probability distribution plots in figure \ref{fig:PDF velocity} for the Aluminium shows the steeper slope of $\sim 5.49$ than in the Tungsten $\sim 4.84$, indicating a lower intermittency in Aluminium than in the Tungsten. 

\begin{figure}[H]
    
    \subfloat{{\includegraphics[width=9 cm]{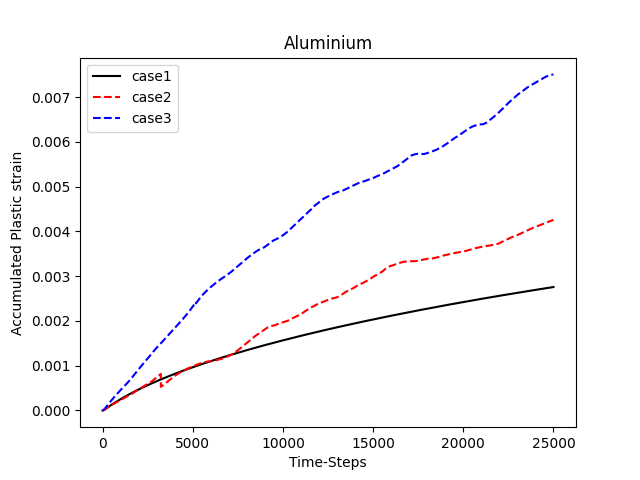} }}
    \subfloat{{\includegraphics[width= 9 cm]{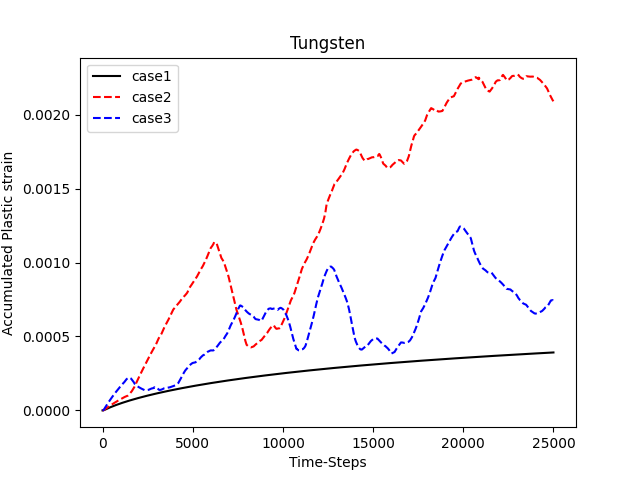} }}
  
    \caption{Accumulated Plastic Strain}%
    \label{fig:accgamma}

\end{figure}

Plastic deformation in the materials can be attributed to the glide of large number of dislocations, number of obstacles,  active slip planes (glide planes), the applied load and many other factors. Ductile materials show the higher plastic deformation than the brittle before failure. In this study, we show that for the same condition, Aluminum shows higher plastic strain than the Tungsten. It might be counter intuitive to assume that the dislocations moving with a higher velocity should posses the higher plastic strain. 
It becomes crucial to consider other factors such as presence of obstacles, and the non-local effect of nearby dislocations also affect the plastic strain of the materials. Figure \ref{fig:accgamma} shows the accumulated plastic strain of each materials for a single realization. Aluminium crystal shows the higher plastic strain value than that of the Tungsten. 

Strain bursts in brittle material are more frequent than in ductile materials. This is due to the  stochastic nature of dislocation movement and the interaction within the crystalline domain. It is evident that brittle materials endure low plastic strain before failure, while ductile materials can withstand significantly high plastic strain. From figure \ref{fig:accgamma}, we observe higher fluctuation of the strain curve in Tungsten than the Aluminium. Which implies that intermittency also occurs in plastic strain of the crystals. 

\begin{figure}[H]
    
    \subfloat{{\includegraphics[width=9 cm]{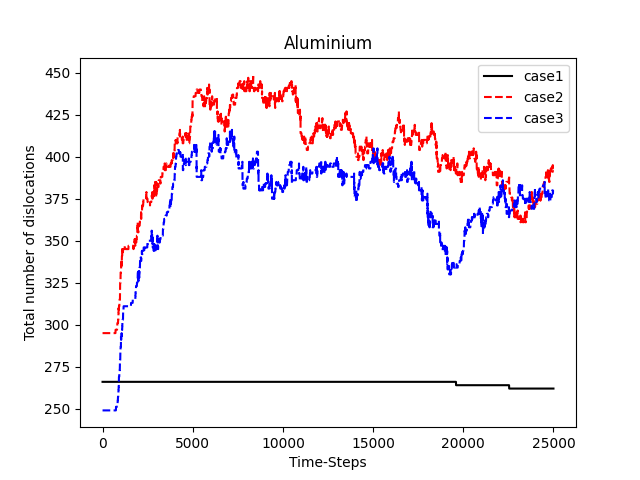} }}
    \subfloat{{\includegraphics[width= 9 cm]{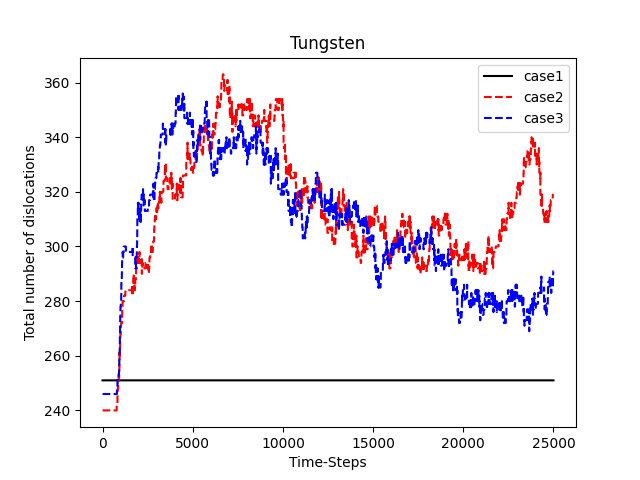} }}
  
    \caption{Total number of dislocation}%
    \label{fig:dislocations}

\end{figure}

The presence of moving dislocations directly influences the accumulated plastic strain that a material can endure. In Figure \ref{fig:dislocations}, we present the evolution of the total number of dislocations. Notably, a greater number of dislocations is observed in the Aluminum compared to Tungsten. Which implies that the higher concentration of moving dislocations in Aluminium contributes to its elevated levels of accumulated plastic strain.

In dislocation dynamics, the formation of junctions occurs when two dislocations of the same sign collide, influencing the dislocation mobility. While simplified 2D-DDD simulations may not fully capture this complex phenomenon \cite{sills2016fundamentals}, an approach inspired by \cite{keralavarma2016strain} is adopted to replicate the junction formation in a simplified manner. The adopted method involves monitoring dislocations within a predefined distance, typically set as \( j_s = 2b \). When two dislocations of the same sign approach within this distance threshold, their positions are fixed or locked. Subsequently, at each time step, the total stress between these locked dislocations is computed, while the mutual stress between them is set to zero. Assuming that the junction breakage is the opposite phenomenon of multiplication, if the stress exceeds a critical threshold, and the elapsed time exceeds a nucleation threshold, the junction is considered to be broken.

Dislocations interactions such as attraction, repulsion, annihilation and junction formations are more common in ductile materials than brittle. It is evident from figure \ref{fig:dislocations} that higher number of dislocations in Aluminium results in higher interactions resulting in higher number of junctions. It may be counter intuitive to consider that having a higher number of junctions hinders the dislocation movements resulting in a less plastic strain. In figure \ref{fig:junctions}, we observe that dislocations in the ductile materials are more active than in brittle materials resulting in higher number of junction formations. 
\begin{figure}[H]
    
    \subfloat{{\includegraphics[width=9 cm]{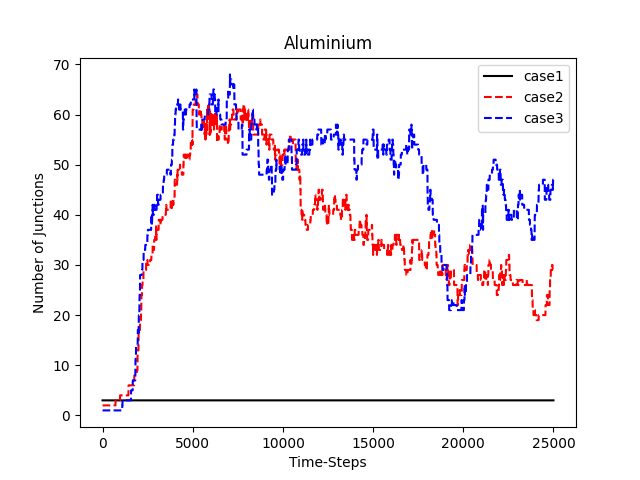} }}
    \subfloat{{\includegraphics[width= 9 cm]{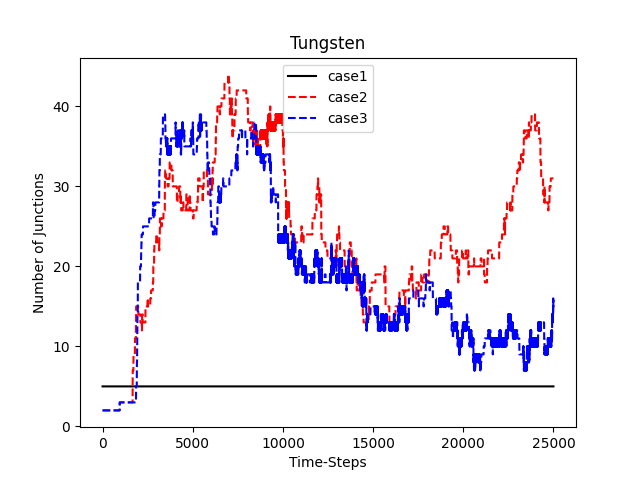} }}
  
    \caption{Number of junctions formed over the time}%
    \label{fig:junctions}

\end{figure}

One approach to understanding dislocation dynamics is to view it through the lens of particle dynamics, where dislocations show stochastic movement. The goal is to analyze the statistics and develop a stochastic process that governs these movements, generating numerous particle trajectories that collectively represent the fluid-like behavior of the system. However, describing dislocation dynamics is challenging. While velocity distributions provide insight into the stochastic nature of the process, it's not straightforward. Some dislocations remain immobile, others exhibit random motion around an equilibrium position, and a few move sporadically with high velocities in a correlated manner. Hence, to capture the fluid-like behavior accurately, it's crucial to gather statistics from a sufficient number of dislocations. The approach for achieving this will be outlined in the following section.

\section{Data Generation}

This section outlines our approach for deriving empirical PDFs directly from DDD simulations. This method eliminates the necessity of constructing a stochastic process, as was previously discussed. While this method incurs higher costs due to the necessity of simulating multiple DDD scenarios, it offers the advantage of utilizing high-fidelity data, thus providing the most accurate depiction of the dynamics attainable.

\subsection{Obtaining Data of Shifted Position}
We begin with introducing the shifted position $X_i(t)$, which is the absolute displacement of $i$ relative to its initial position. It's calculated by comparing the initial position $x_i(0)$, with the current position, denoted as $x_i(t)$,

\begin{equation}
X_i(t) = x_i(t) - x_i(0),\quad \text{for} \quad t \in (0,T].
\end{equation}

We gather enough dataset, denoted as $X_i(t)$, by treating the Dislocation Dynamics (DDD) simulation as a stochastic process. We perform 2000 DDD simulations, each initialized with randomly distributed dislocations and frank read sources. To efficiently handle the simulations, we distribute the computations across multiple HPC cores, exploiting parallelism. With the number of dislocations typically ranging between 200 and 300 after relaxation, aggregating the shifted positions of all dislocations across the 2000 simulations yields trajectories for approximately 730K for the aluminium and 640K for tungsten,  Lagrangian particles. These particles are stochastic in nature, originating from $X_i(0) = 0$.

We transform the trajectories of Lagrangian particles obtained from DDD simulations into a time-varying PDF denoted as $\hat{p}(x,t)$. This PDF indicates the likelihood of finding a dislocation at a distance $x$ from its starting position since the beginning DDD simulation. Our objective is to formulate a model for the evolution of $\hat{p}(x,t)$, which we hypothesize is governed by an integral operator similar to a nonlocal Laplacian. In the following section, we will discuss the procedure for converting DDD position data into density estimates, which will be the training datasets for our ML algorithm.

\subsection{Density Estimation}
In our proposed nonlocal model for the evolution of the dislocation position PDF, our primary interest lies in understanding the dynamics of dislocation particles under constant load, coupled with the constantly changing stress landscape resulting from the change is dislocation density (addition and removal of other dislocations). We focus on how dislocations respond to load, multiplication, annihilation, and junction formation mechanisms in a broader context. In this regard, when considering an infinite number of particles and aiming to capture their collective behavior without explicitly modeling creation and destruction events, we omit birth and death processes from the nonlocal formulation. Instead, our focus is solely on describing the nature of dislocation motion from a continuum perspective, which emerges as a consequence of these mechanisms observed at the discrete level of representation in DDD level.

\begin{figure}[H]
    
    \subfloat{{\includegraphics[width=9 cm]{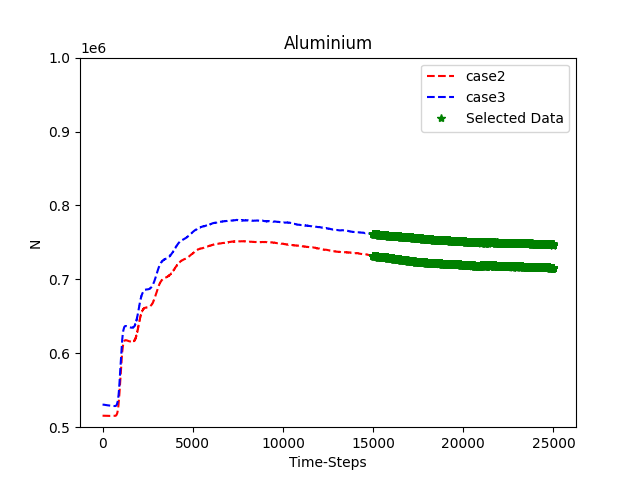} }}
    \subfloat{{\includegraphics[width= 9 cm]{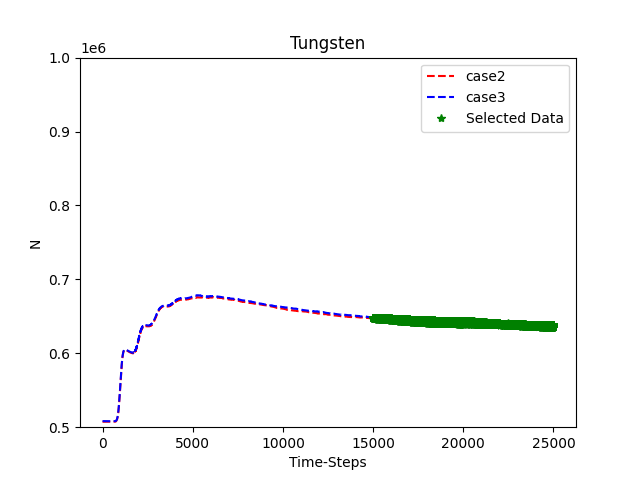} }}
  
    \caption{Total number of dislocations during nr = 2000 realization for both cases. Bold green line indicates the selected data points for the ML algorithm.}%
    \label{fig:N Dislocations }

\end{figure}

In the context of DDD, the creation and annihilation of dislocations introduce noise when representing the system as a continuous PDF. To minimize this noise, especially in the creep regime, we select a time frame in which the total number of dislocations are almost constant, or change by a very little fraction. For training and testing our ML framework, we choose the last 10000 time steps from the DDD simulations. This selection ensures a dataset with minimal fluctuations in the dislocation density during the DDD simulation. In this chosen time series, we observe only negligible relative differences in the number of dislocations between the initial and final states. Figure \ref{fig:N Dislocations } shows the selected data points from the DDD simulation for further processes. In this figure, we observe that number of dislocations increases with increasing load in case of Aluminium. However, the total number of dislocations for Tungsten is almost same regardless of the external load. This implies that the dislocations in ductile materials are more sensitive to the applied load. 

\subsubsection{Kernel Density Estimation (KDE)}

Kernel Density Estimation (KDE) is a non-parametric method that doesn't rely on assumptions about the underlying distribution. However, KDE requires the selection of a specific kernel function $k(x-x_i;w_0)$, which is determined by a bandwidth parameter $w_0$. Initially, we obtain a density estimate $\hat{p}_0(x)$ through this process.

\begin{equation}
\hat{p}_0(x) = \frac{1}{n}\sum_{i=1}^n k(x-x_i;w_0),
\end{equation}

\noindent where $x$ is the coordinate on which we are aiming to calculate PDF, and $x_i$ are locations $i = 1,\,2,\,\dots,\,n$.

The kernels are normalized to unity, i.e.,
\begin{equation}
\int_{-\infty}^{\infty}k(x;w_0)dx = 1,
\end{equation}

\noindent and take the form

\begin{equation}
k(x-x_i;w_0) = \frac{1}{w_0} K\left(\frac{x-x_i}{w_0}\right).
\end{equation} 

Then, the final density estimator will be;

\begin{equation}
\hat{p}_0(x) = \frac{1}{nw_0}\sum_{i=1}^n K\left(\frac{x-x_i}{w_0}\right).
\end{equation}

When the number of data points $n$ tends towards infinity for normally distributed data, the Mean Integrated Square Error of $\hat{p}_0$ is minimized according to \cite{silverman2018density}

\begin{equation}
w_0 = 1.06 s n^{-1/5}.
\label{nlml_eq:w0}
\end{equation}

\noindent where s is the standard deviation of the sample

\subsubsection{Adaptive Kernel Density Estimation (AKDE)}

Due to the significant jumps observed in DDD, we anticipate that the PDFs will exhibit heavy tails. However, with limited data, the majority of the distribution's mass will likely concentrate in the central region. Employing a uniform binning method like KDE could introduce noise in the tails, which is undesirable as these curves serve as input for the ML algorithm. To address this issue, we utilize AKDE to derive a continuous, smooth function.

We initiate the AKDE process using an initial classical KDE estimate $\hat{p}_0(x)$ with a constant bandwidth $w_0$ determined by Eq.~(\ref{nlml_eq:w0}). Subsequently, adaptivity occurs in the bandwidth of each data point \cite{silverman2018density}.

\begin{equation}
w_i = w_0 \lambda_i = w_0 \left( \frac{\hat{p}_0(X_i)}{G}\right)^{-\xi},
\end{equation}

\noindent where $G$ is,

\begin{equation}
G = \exp \left(\frac{1}{n} \sum_{i=1}^n \ln \hat{p}_0(X_i)\right).
\end{equation}

Additionally, a sensitivity parameter $0 \leq \xi \leq 1$ regulates the significance of the initial guess's shape compared to the subsequent estimation \cite{pedretti2013automatic}. Theoretically, an optimal value for $\xi$ was identified as $\xi = 0.5$ \cite{abramson1982bandwidth}. Ultimately, the density from AKDE is derived as

\begin{equation}
\hat{p}_1(x) = \frac{1}{n} \sum_{i=1}^n \frac{1}{w_i} K\left(\frac{x-x_i}{w_i}\right).
\end{equation}

We define a support domain $\Omega$ for the density function by providing extra zeros beyond the outermost points in the $x$ direction. This extension forms a buffer zone between the nonlocal simulation domain and compact support of the PDF. Then, $\hat{p}_1$ are then computed at regularly spaced points within $\Omega$, using a constant grid of $n = 1601$ points for both crystals. For aluminum, $\Omega_2$ and $\Omega_3$ range from -65 to 65, whereas for tungsten, both $\Omega_2$ and $\Omega_3$ range from -45 to 45. The computation of $\hat{p}_1(x)$ at each time step is parallelized, with each core handling a distinct time step, involving a total of 1000 HPC cores for each case. Finally, after obtaining the final estimates $\hat{p}_1(x)$ for all time steps, a symmetrization step is performed to ensure consistency in the ML algorithm. This step is crucial because a symmetric, radial-basis nonlocal kernel is used in defining our operator. The symmetrization process involves monitoring the evolution of the mean $\mu$ and the skewness factor $\tilde{\mu}_3$ from $\hat{p}_1(x)$. This is defined as

\begin{equation}
\mu = \mathbb{E}[\hat{p}_1(x)] = \int \hat{p}_1(x) x  dx,
\end{equation}

\begin{equation}
\tilde{\mu}_3 = \mathbb{E}\left[\left(\frac{\hat{p}_1(x)- \mu}{\sigma}\right)^3\right] = \frac{\int (x-\mu)^3 \hat{p}_1(x) dx}{\left( \sqrt{\int (x-\mu)^2 \hat{p}_1(x) dx}\right)^3}.
\end{equation}

\begin{figure}[H]
	\centering
	\subfloat[]{\includegraphics[width=0.24\textwidth]{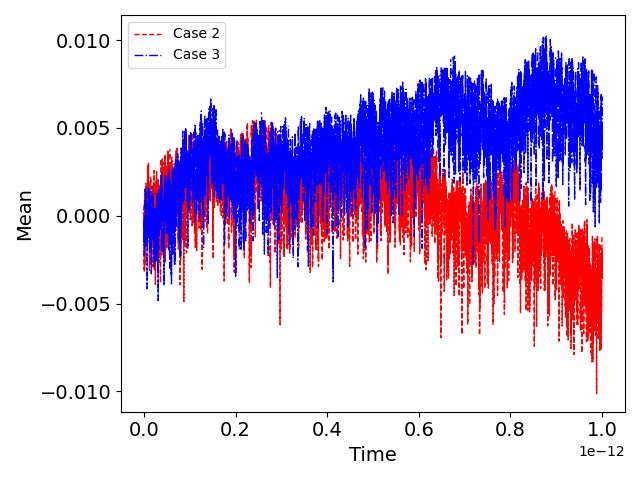}}
	\subfloat[]{\includegraphics[width=0.24\textwidth]{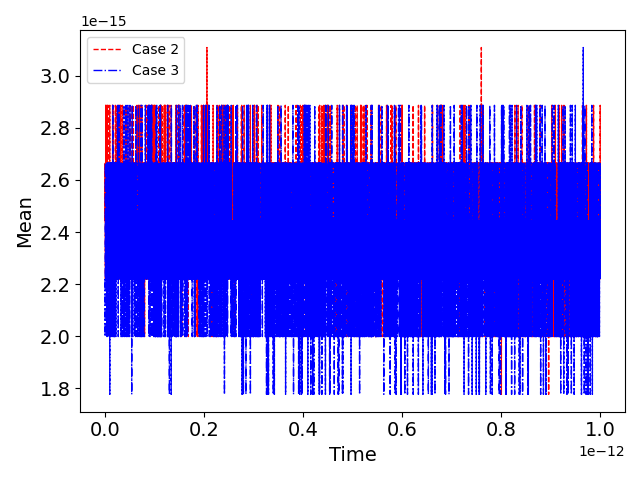}}
	\subfloat[]{\includegraphics[width=0.24\textwidth]{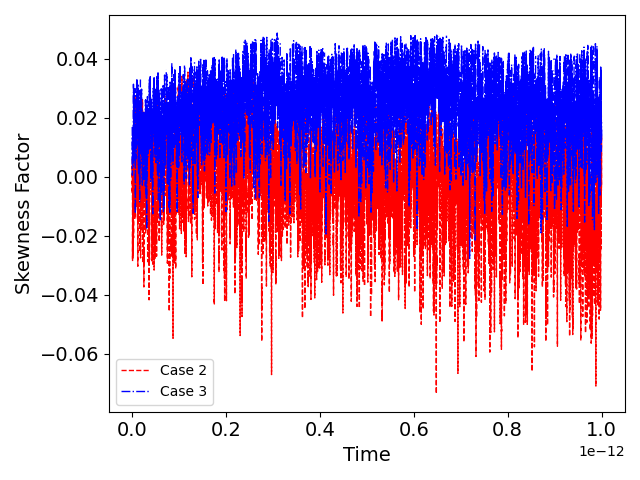}}
	\subfloat[]{\includegraphics[width=0.24\textwidth]{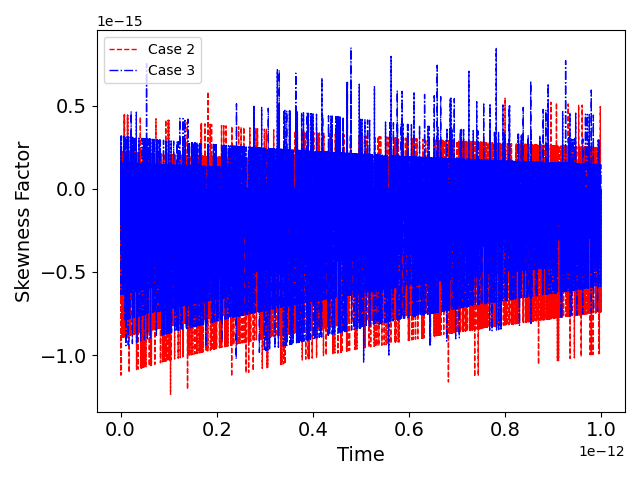}}
    \hfill
    \subfloat[]{\includegraphics[width=0.24\textwidth]{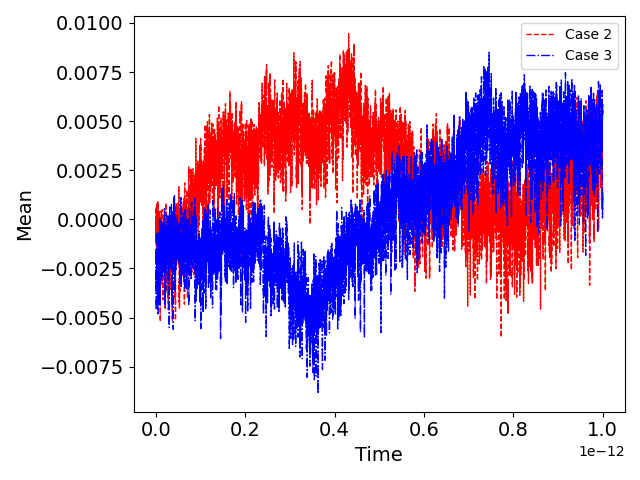}}
	\subfloat[]{\includegraphics[width=0.24\textwidth]{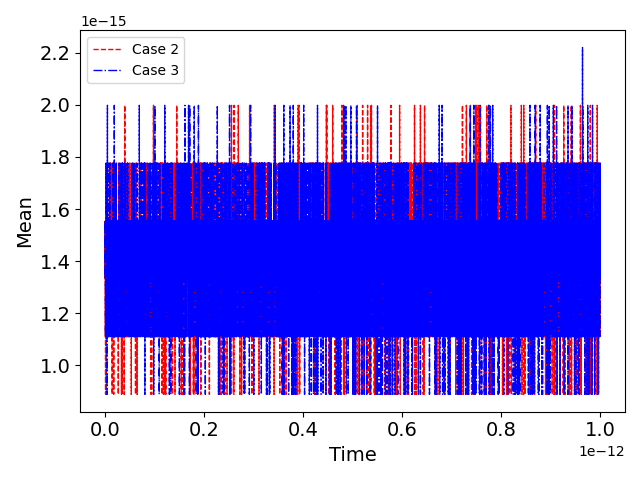}}
	\subfloat[]{\includegraphics[width=0.24\textwidth]{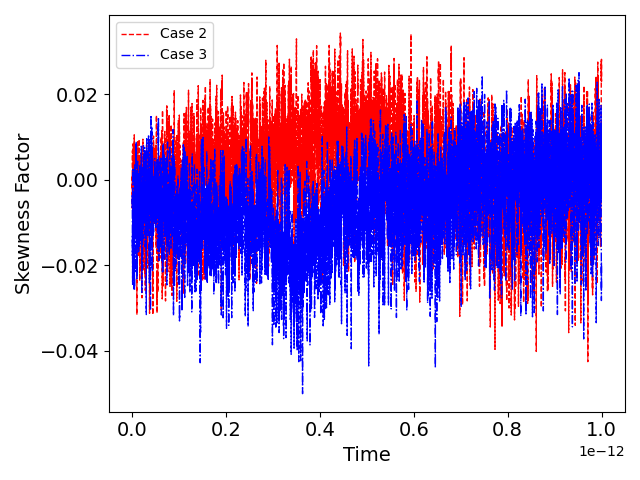}}
	\subfloat[]{\includegraphics[width=0.24\textwidth]{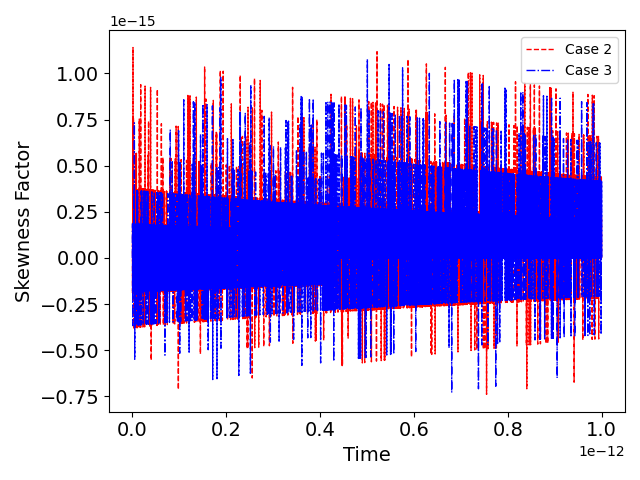}}
	\caption{Time evolution of mean and skewness factor before and after  symmetrization and re-normalization. (a)-(d) for the Aluminium, and (e)-(h) for the Tungsten.}
	\label{fig:skew-mu}
\end{figure}

First, we ensure that any nonzero values of these parameters are minimal, attributing them to insufficient data. Then, we mirror the left side of the function $\hat{p}_1(x)$ to the right side and adjust the probability density function (PDF) accordingly for symmetry. This ensures a more suitable fit for the radial-basis nonlocal kernel. Figure \ref{fig:skew-mu} illustrates the changes in the values of 
$\mu(t)$ and  $\tilde{\mu}_3(t)$  before and after this symmetrization process.

\begin{figure}[H]
	\centering
	\subfloat[case2.]{\includegraphics[width=0.48\textwidth]{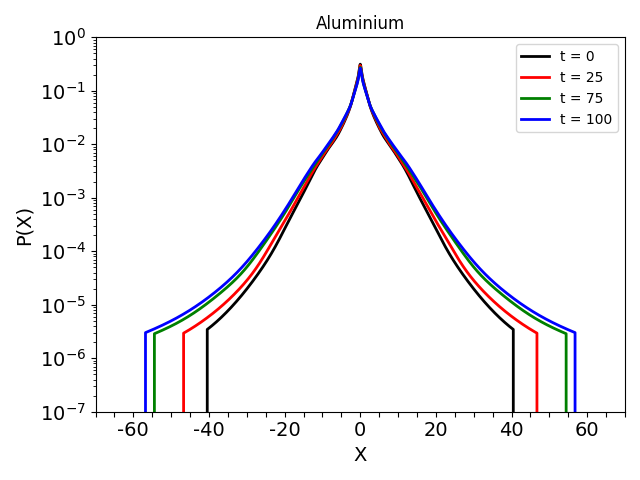}}
	\subfloat[case3.]{\includegraphics[width=0.48\textwidth]{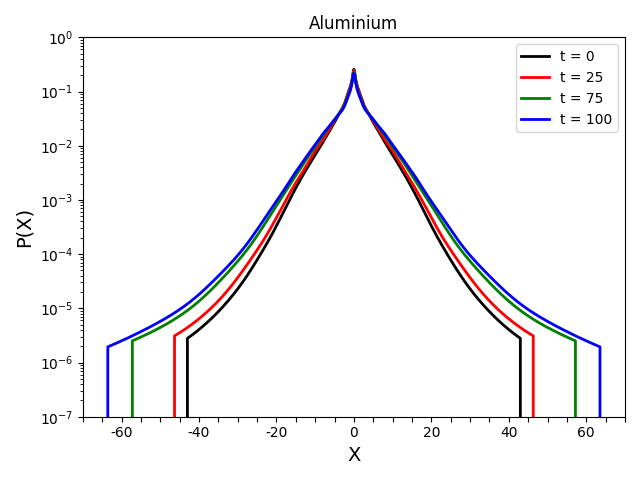}}
    \hfill
	\subfloat[case2.]{\includegraphics[width=0.48\textwidth]{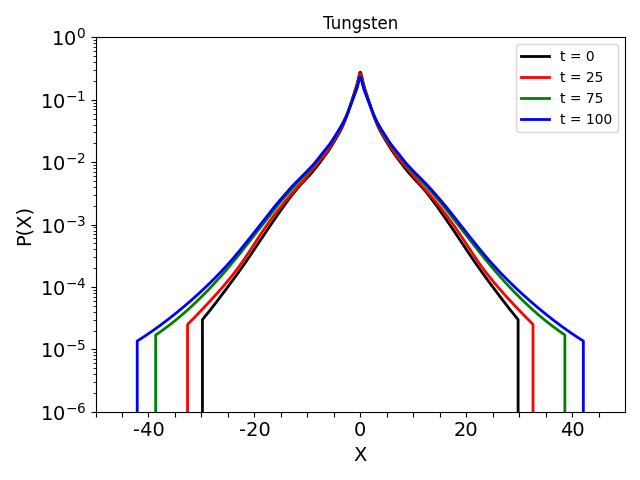}}
	\subfloat[case3.]{\includegraphics[width=0.48\textwidth]{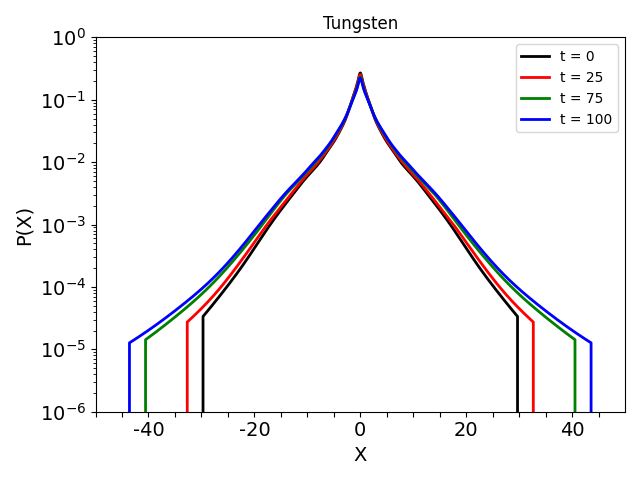}}
    
	\caption{PDF plots of scaled shifted position. (a)-(b) for the Aluminium, and (c)-(d) for the Tungsten}
	\label{fig:pdfs}
\end{figure}

In Figure \ref{fig:pdfs}, we plot the PDF of the scaled shifted position $X_t$ of the dislocations for both aluminium and tungsten. The scaled shifted position is defined as the ratio of the absolute displacement $x_t$ to the respective Burgers' vector $(b)$. Both materials exhibit power-law tails in their PDFs. We observe that the support domain for aluminium is significantly higher than that for tungsten, implying that ductile materials tend to have larger support domains for long-range interactions. For instance, the support domain for \textbf{case 2} in aluminium is approximately 58, whereas for tungsten it is around 42. Additionally, the externally applied load has a significant effect on ductile materials. This is evident as the support domains for \textbf{case 2} and \textbf{case 3} are different for aluminium. However, the same change in applied load has a lesser effect on brittle materials like tungsten.

\subsection{Nonlocal Transport Models}

To monitor the progression of PDFs describing dislocation positions, we propose the use of a parabolic nonlocal transport model. This model is governed by a nonlocal operator that incorporates a kernel function, the specifics of which must be determined. To achieve this, we utilize data generated from DDD simulations. By employing this simulation data, we train a machine learning surrogate model that is capable of identifying and optimizing the parameters of our proposed transport model.

We denote the empirical PDF estimate at a given time $t$ by $p(x,t)$. This PDF is considered within the temporal interval $[0, T]$, where the variable $x$ signifies the position over the spatial domain $\Omega = [-L, L]$. The parameter $L$ is determined by extending the maximum support observed at the final time-step of the simulation, and then adding additional zeros to accommodate this extension. This process is depicted in Fig.~\ref{fig:pdfs}. The time evolution of the function $p$ is described by a nonlocal parabolic equation, which governs the changes in the PDF over time.

\begin{equation}
\begin{cases}
\dot{p}(x,t) = \mathcal{L}p, & x \in \Omega \\
\mathcal{B}_{\mathcal{I}}p(x) = g(x), & x \in \Omega_{\mathcal{I}},
\end{cases}
\label{nlml_eq:nonlocal_evolution}
\end{equation}

\noindent where $\mathcal{L}$ denotes the nonlocal (linear) Laplacian operator defined as

\begin{equation}
\mathcal{L}p = \int_{B_\delta(x)} K(\vert y - x \vert) (p(y) - p(x)) dy.
\label{nlml_eq:nonlocal_laplacian}
\end{equation}

\( B_\delta(x) \) denotes the ball centered at \( x \) with radius \( \delta \), commonly referred to as the \textit{horizon}, which determines the compact support of \( \mathcal{L} \). It is significant to mention that for certain kernel functions, \( \mathcal{L} \) aligns with well-known operators, such as the fractional Laplacian \cite{gulian2021unified, d2013fractional}. Specifically, when the kernel function is \( K(|y-x|) \propto |y-x|^{-\alpha} \), with \( \alpha=1+2s \) and \( s \in (0,1) \), the operator \( \mathcal{L} \) is equivalent to the one-dimensional fractional Laplacian. Additionally, when this kernel is limited to the compact support \( B_\delta(x) \), \( \mathcal{L} \) corresponds to the truncated fractional Laplacian, which is the preferred operator in our approach.

The region where nonlocal boundary conditions (or volume constraints) are specified is defined as follows:

\begin{equation}
 \Omega_{\mathcal{I}} = \{ y \in \mathbb{R}\setminus\Omega \text{ such that } \vert y - x\vert < \delta \text{ for some }x \in \Omega\}.
\end{equation}

We impose nonlocal homogeneous Dirichlet volume constraints, described in one dimension by the nonlocal interaction operator \( B_\mathcal{I}:[-L-\delta,-L) \cup (L,L+\delta] \to \mathbb{R} \), ensuring that \( g(x) = 0 \) for \( x \in \Omega_\mathcal{I} \).

\section{Nonlocal Power-Law Parameters for Ductile and Brittle Materials}

In our previous work \cite{de2023machine}, we developed a ML framework to address the inverse problem of determining the parameters of a nonlocal equation using high-fidelity data. In particular, we input PDFs derived from DDD simulations into the ML algorithm to retrieve the parameters of the nonlocal power-law kernel, specifically the fractional order \( \alpha \), horizon \( \delta \), and a linear coefficient. We split our data into train (80$\%$) and test (20$\%$) datasets, and feed the data into our developed ML model and obtained the nonlocal parameters for both materials. Table \ref{tab:Parameters} contains the obtained values for both materials for each cases.

\begin{table}[h]
\centering
\caption{Nonlocal parameters of Aluminium and Tungsten}
\begin{tabular}{ccccc}
\hline
& \multicolumn{2}{c}{Aluminium} & \multicolumn{2}{c}{Tungsten} \\ \hline
& $\alpha$ & $\delta$ & $\alpha$ & $\delta$ \\ \hline
\textbf{Case2} &2.15 &52.53 &1.96 &36.30 \\ 
\textbf{Case3} &2.14 &54.55 &1.96& 37.14\\ \hline
\end{tabular}

\label{tab:Parameters}
\end{table}

We start with the inital guess of $\alpha_{true}  = 2$, and $\delta_{true} = 0.5L$, for both materials of each cases. From the ML algorithm, we obtain the fractional order $(\alpha)$ to be 2.15 for the Aluminium.  Similarly we obtain $\alpha$ to be 1.96 for the Tungsten. This implies that the dislocation motions in ductile materials are mixed anomalous and diffusive/hyper-diffusive while brittle materials are super-diffusive in nature.

\begin{figure}[H]
	\centering
	\subfloat[Al-case2.]{\includegraphics[width=0.48\textwidth]{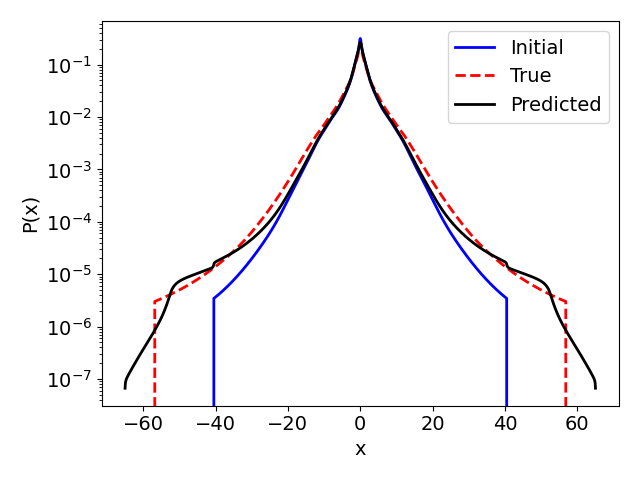}}
	\subfloat[Al-case3.]{\includegraphics[width=0.48\textwidth]{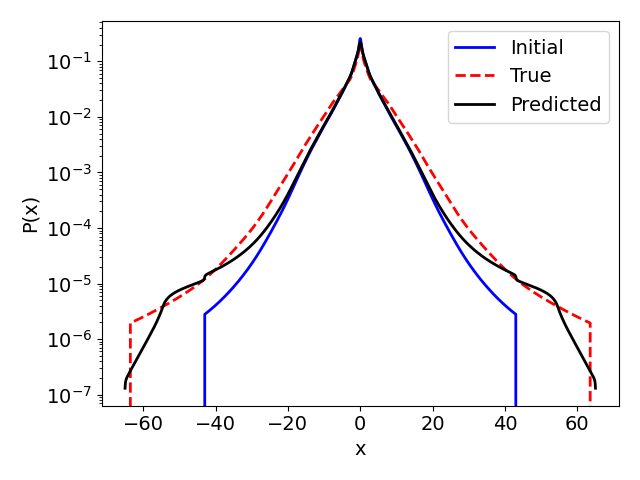}}
    \hfill
	\subfloat[W-case2.]{\includegraphics[width=0.48\textwidth]{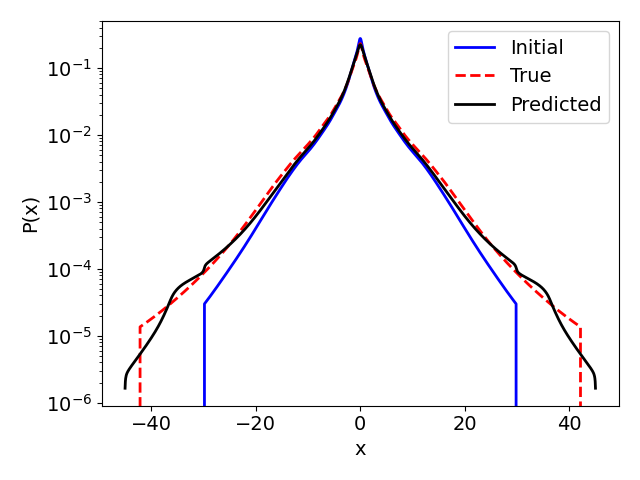}}
	\subfloat[W-case3.]{\includegraphics[width=0.48\textwidth]{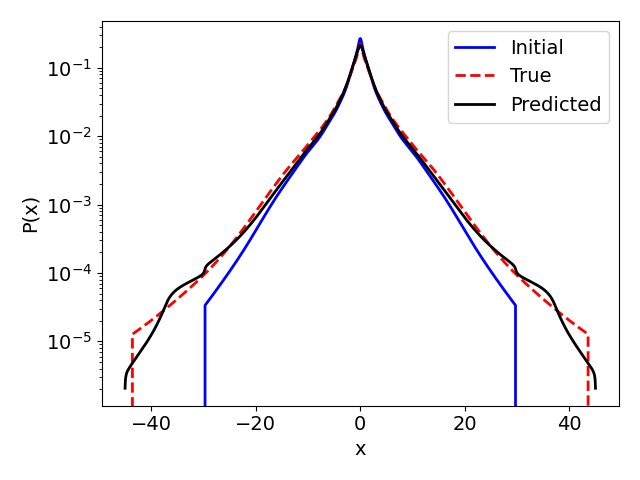}}
    
	\caption{Simulation of the nonlocal model for the whole time-interval of available data, highlighting the initial PDF, and the final distributions of the true data and the nonlocal prediction. (a)-(b) for the Aluminium, and (c)-(d) for the Tungsten}
	\label{fig:prediction}
\end{figure}

Finally, we use optimal parameters and feed into our proposed nonlocal model and run the simulation. In figure \ref{fig:prediction}, we observe that obtained optimal parameters can reproduce the original PDF. For the case of Aluminium, we observe that initially, the shape follows the initial PDF later it diverges towards the true shape with relative error of $\epsilon_2 = 0.0888 $ for case2 and $\epsilon_3 = 0.1315$ for case3. On the other hand, for Tungsten, the predicted PDF shape follows the true shape all the way to final timestep with relative error of $\epsilon_2 = 0.0606 $ for case 2 and $\epsilon_3 = 0.0612$ for case 3.  

\section{Conclusion}
 
Under similar external conditions, dislocations behave differently in brittle and ductile materials. In brittle materials, dislocations exhibit more intermittent behavior and tend to move with higher velocities than in ductile materials. However, despite their higher velocities, these dislocations result in less plastic strain. This suggests that dislocations in brittle materials form clusters and remain close to their initial positions. Conversely, in ductile materials, dislocations encounter less resistance, allowing them to travel farther from their initial positions. Velocity and strain plots further reveal that strain bursts are also intermittent in brittle materials. Furthermore, natural phenomena such as dislocation multiplication and junction formation is more frequent in ductile materials than brittle. We have observed higher dislocation density in figure \ref{fig:dislocations}, junctions formation in figure \ref{fig:junctions}, in addition to that, figure \ref{fig:N Dislocations } shows that total number of dislocations in overall DDD simulation are higher for ductile materials.

We proposed a nonlocal transport model governed by a nonlocal operator with a kernel function applicable to both materials. Using high-fidelity data from DDD simulations, we fed our developed model without altering input parameters. Our findings indicate that the model, developed from non-dimensional data, can be effectively applied to any specific material case. The operator $\mathcal{L}$, derived from a nonlocal power-law kernel with a fixed horizon, is truncated fractional Laplacian of fractional order $s$ in the form $\alpha = 1 + 2s$. For aluminum, we identified the fractional order $s \approx 0.57$, and for tungsten, $s \approx 0.48$. These values suggest that dislocation motion in ductile materials exhibits a mix of anomalous and diffusive/hyper-diffusive behavior, while in brittle materials, the motion is predominantly super-diffusive.

\section{Data availability}

Data can be provided upon request


\bibliographystyle{ieeetr}
\bibliography{references}
\end{document}